\documentclass[conference]{IEEEtran}

\usepackage[T1]{fontenc}% optional T1 font encoding

\ifCLASSINFOpdf
\else
\fi
 \usepackage{graphicx}
          \usepackage{float}

\usepackage{amsmath}
\usepackage[cmintegrals]{newtxmath}
\begin{document}
\title{Analysis of Full-Duplex Downlink Using Diversity Gain}

\author{\IEEEauthorblockN{Chandan Pradhan and Garimella Rama Murthy}
\IEEEauthorblockA{Signal Processing and Communication Research Center\\
IIIT Hyderabad, India\\
Email: chandan.pradhan@research.iiit.ac.in, rammurthy.iiit.ac.in}}


%{Shell \MakeLowercase{\textit{et al.}}: Bare Demo of IEEEtran.cls for IEEE Communications Society Journals}

\markboth{IEEE SIGNAL PROCESSING LETTERS}%
{Shell \MakeLowercase{\textit{et al.}}: Bare Demo of IEEEtran.cls for IEEE Communications Society Journals}




\maketitle

\begin{abstract}
The paper carries out performance analysis of a multiuser full-duplex (FD) communication system. Multiple FD UEs share the same spectrum resources, simultaneously, at both the uplink and downlink.  This results in co-channel interference (CCI) at the downlink of a UE from uplink signals of other UEs. This work proposes the use of diversity gain at the receiver to mitigate the effects of the CCI. For this an architecture for the FD eNB and FD UE is proposed and corresponding downlink operation is described. Finally, the performance of the system is studied in terms of downlink capacity of a UE. It is shown that through the  deployment of sufficient number of transmit and receive antennas at the eNB and UEs, respectively, significant improvement in  performance can be achieved in the presence of CCI.  
\end{abstract}


% \fi
\IEEEpeerreviewmaketitle

\section{Introduction}

\IEEEPARstart{T}{he} inevitable high bandwidth requirement in the future cellular network has made researchers to come up with revolutionary ideas in recent times. One such idea is the introduction of full-duplex (FD) communication. A FD systems make the simultaneous in-band transceiving feasible, i.e, simultaneous uplink and downlink operation using the same spectrum resources. In recent years, extensive work has been done in the area of self-interference cancellation (SIC) design, including for compact devices like laptops and smart phones, enabling FD communication for both single and multiple antenna transceiver units \cite{full_duplex, compact}. The designs aim in optimal cancellation of interference from the receiver chains introduced by the transmitter chains of the transceiver unit. While this is far from true today for cellular networks, sufficient progress is being made in this direction to start considering the FD model and its implications \cite{fd_small}. \par

     In the conventional LTE system, for uplink and downlink, single carrier frequency division multiple access (SC-FDMA) and orthogonal frequency division multiple access (OFDMA) is used for multiple access respectively. For FD operation, same subcarriers can be allocated to a UE for uplink and downlink. Hence, the use of SC-FDMA for both uplink and downlink is proposed \cite{ants}, due to its advantages over OFDMA in terms of bit error rate (BER) performance and energy efficiency \cite{sc_fdma}. \par
     
    In this work, multiple user equipments (UEs) are considered operating in FD mode on the same set of subcarriers simultaneously. However, the use of the same subcarriers for both uplink and downlink results in co-channel interference (CCI) in downlink of a UE from uplink signals of other UEs operating in the same subcarriers. The prior work \cite{ants}, considers deploying the smart antenna technique at the UEs with highly spatially correlated multiple antennas.  In this work, a converse scenario is considered with a highly scattered environment which prevents the use of the method proposed in \cite{ants} to mitigate the CCI. Hence, here the use of diversity gain at the UEs is analyzed to mitigate the effect of CCI and allowing the UEs to share the same spectrum resources. Also, the proposed architecture for the UE has less computational complexity than the architecture deploying smart antenna technology proposed in the previous work. \par
    
    The rest of the paper is organized as follows. In section 2, the system model for the proposed method is discussed. Section 3 presents the downlink operation and diversity gain methodology to tackle the CCI at the downlink of a UE. The system performance is analyzed through simulations in section 4. Finally, we conclude in section 5. \par
    
    Notation: $[.]^T$,$(.)^H$ denote transpose and Hermitian respectively. $||.||$  denotes the Euclidean norm. $(.)^d$ and $(.)^u$ denote downlink and uplink components, respectively. \par
    
\section{System Model}

In this work, for facilitating FD communication, both the eNB and UEs operate in FD mode. An FD eNB with $N_e$ antennas and $K$ FD UEs with $N_r$ antennas is considered each such that $N_e \geq KN_r$. Assuming highly scattered environment, the number of data streams per UE is given by $Q=N_r$. For the proposed transceiver architecture shown in  fig.1 and fig.2 for eNB and UE respectively, the Analog and Digital SIC unit at the RF front end,  includes the SIC circuitry  \cite{full_duplex,compact} enabling the FD communication.  The details of SIC design are not discussed here. \par    

 \begin{figure}
\centering
\includegraphics[width=3.25in ,height=2.0in]{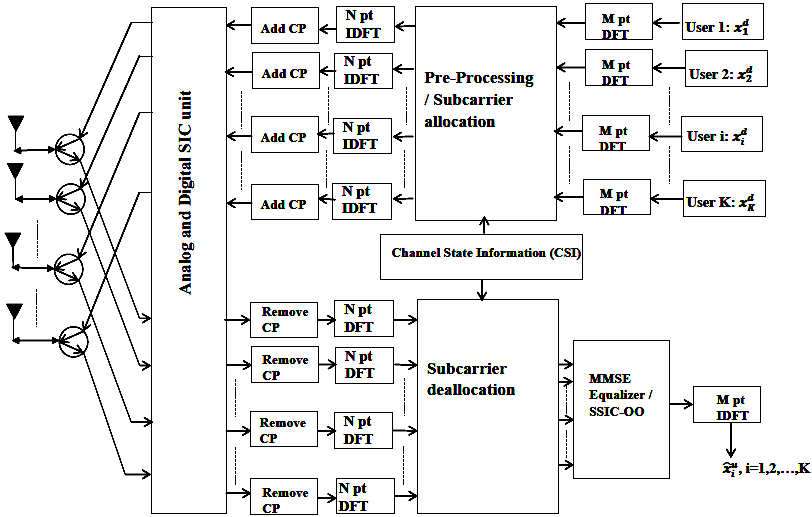}
  % where an .eps filename suffix will be assumed under latex, 
% and a .pdf suffix will be assumed for pdflatex; or what has been declared
% via \DeclareGraphicsExtensions.
\vspace{-0.5em}
   \caption{Transceiver structure for the proposed eNB architecture}
   \label{fone}
\end{figure}

\begin{figure}
\centering
\includegraphics[width=3.25in ,height=1.75in]{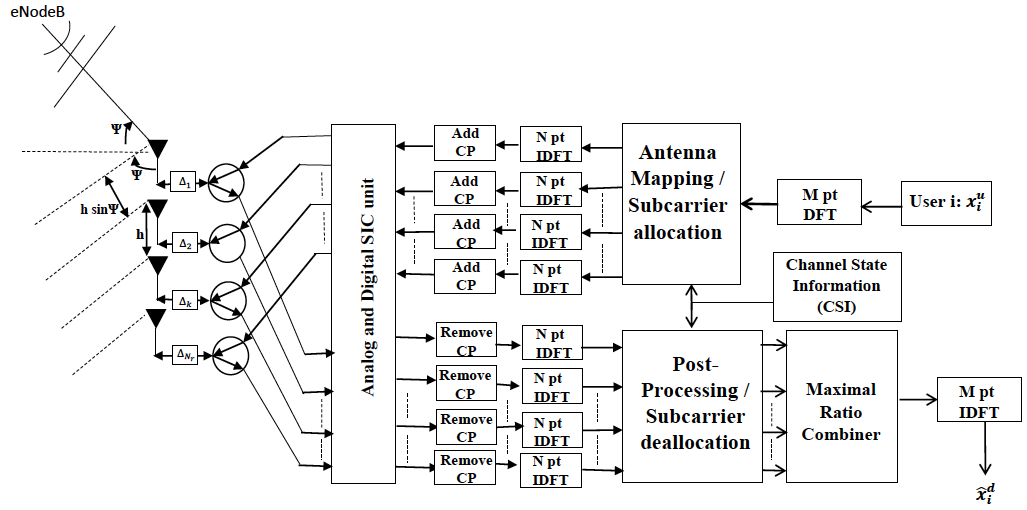}
  % where an .eps filename suffix will be assumed under latex, 
% and a .pdf suffix will be assumed for pdflatex; or what has been declared
% via \DeclareGraphicsExtensions.
\vspace{-0.5em}
   \caption{Transceiver structure for the proposed UE architecture}
   \label{ftwo}
   \vspace{-1.5em}
\end{figure}

  Let each subcarrier allocated be shared by $\acute{K}$ UEs simultaneously, where $\acute{K}$ is given by \cite{sc_fdma}:
  
  \begin{equation}
      \label{one}
     \acute{K} = min\Big( \left \lfloor{\frac{N_e}{N_r}}\right \rfloor ,K \Big) 
\end{equation}. 

Each UE is allocated M (=$\left \lfloor{\frac{N\acute{K}}{K}}\right \rfloor$ ) subcarriers, where $N$ is the total number of subcarriers available.  Keeping this in mind, a case of $\acute{K}=K$, i.e., all the $K$ UEs are allocated all the $N$ subcarriers, is considered. However, the appropriate number of co-existing UEs depends on the CCI experienced by the UEs in their downlink and hence can be $\leq \acute{K}$. The channel between each eNB antenna and UEs antenna is assumed to be frequency selective with $L$ taps. The FD operation allows the channel reciprocity between downlink and uplink, $h^{u}_{j,i,k} (b) = h^{d}_{j,i,k} (b)$, where  $ h^{u}_{j,i,k} (b)$ and $h^{d}_{j,i,k} (b) $  denotes $b^{th}$ time domain  uplink and downlink channel coefficient between $j^{th}$ antenna at the eNB and $k^{th}$  antenna of the $i^{th}$ UE, respectively, $b = 0, 1, 2, ... , L-1$, $j = 1 , 2, ..., N_e$, $k = 1, 2,...,N_r $ and $i = 1,2, ...,K$. \par

     In downlink, the channel reciprocity property of FD communication enables the transmitter (eNB) to acquire CSI with ease. The CSI can be used for precoding the UE data at eNB to perform the SVD based beamforming. In the uplink, successive interference cancellation with optimal ordering (SSIC-OO) algorithm \cite{ants} is used at the eNB to segregate signals of UEs sharing the same subcarriers. In \cite{ants}, a smart antenna approach is deployed to avoid the CCI at downlink of UEs. Here, the paper concentrates on the \textit{downlink} operation and to avoid CCI at downlink of a UE, diversity combining technique is used to minimize interference from the uplink signals of other co-existing UEs. \par

%\begin{figure}
%\centering
%\includegraphics[width=2.5in ,height=1.5in]{fig3.png}
  % where an .eps filename suffix will be assumed under latex, 
% and a .pdf suffix will be assumed for pdflatex; or what has been declared
% via \DeclareGraphicsExtensions.
%\vspace{-0.5em}
 %  \caption{Structure for the Beamforming unit for the proposed UE architecture and the formation of directed beam toward eNB}
  % \label{fthree}
   %\vspace{-1.0em}
%\end{figure}

\section{Full-Duplex Multiuser Downlink Operation}
    In the downlink, let $\mathbf{x}^d_i$  denote the $i^{th}$ UE information data block of length $M$, shown in fig.\ref{fone}. The output of the M-point DFT block is given by:

\begin{equation}
\label{two}
\mathbf{\bar{x}}_i^d = \mathbf{F}_M \mathbf{x}_i^d
       \end{equation}

where $i=1,2,...,K$, $\mathbf{F}_M$  is the M-point DFT matrix and $\mathbf{x}_i^d = [x_i^d(1),x_i^d(2),...,x_i^d(m),...,x_i^d(M)]^T $. $x_i^d(m)$ is the $m^{th}$ data symbol of $i^{th}$ user and unlike in \cite{sc_fdma} same $M$ symbols are transmitted in all the $Q$ data streams. This output is then passed through the pre-processing / subcarrier allocation block. Let  $\mathbf{P}^i_m$ denotes $N_e X Q$  precoding vector for the $i^{th}$ UE on the $m^{th}$ subcarrier. The precoded output vector of size $N_e X 1$ for the $i^{th}$ UE on the $m^{th}$ subcarrier is given by:

\begin{equation}
\label{three}
\mathbf{z}^d_i(m) = \mathbf{P}^i_m \mathbf{\tilde{x}}_i^d (m)
       \end{equation}   
       
where $m=1,2,...,M$ and $\mathbf{z}^d_i(m) = [z^d_{i,1}(m), z^d_{i,2}(m),...,z^d_{i,N_e}(m)]^T$. $\mathbf{\tilde{x}}_i^d (m)= (\bar{x}_i^d(m) \mathbf{J}_{1,Q})^T$ where $\bar{x}_i^d(m)$ is the $m^{th}$ element of $\mathbf{\bar{x}}_i^d$ and $\mathbf{J}_{1,Q}$ is the $1 X Q$ unit matrix. Let $\mathbf{A}^i$, represents the $N X M$ subcarrier allocation matrix for the $i^{th}$ UE \cite{sc_fdma}. The transmit signal vector, at the $j^{th}$ antenna of the eNB, after subcarrier allocation and IDFT operation is given by:  

\begin{equation}
\label{four}
\mathbf{s}^d_j = \mathbf{\bar{F}}_N \mathbf{e}^d_j
\vspace{-0.10em}
       \end{equation} 
       
where $j=1,2,...,N_e$, $\mathbf{\bar{F}}_N$ denotes the N-point IDFT matrix, $\mathbf{e}^d_j = \sum_{i=1}^{K} \mathbf{A}^i \mathbf{z}^d_{i,j}$ and $\mathbf{z}^d_{i,j}=[z^d_{i,j}(1), z^d_{i,j}(2),...,z^d_{i,j}(M)]^T$. This signal is then transmitted after the addition of the cyclic prefix (CP). \par

In this work, the UE transceiver unit (Fig.\ref{ftwo}) consists of a uniformly spaced linear antenna array of $N_r$ elements with an inter element distance of $h$. The angle with respect to the array normal at which the plane wave impinges upon the array is represented as $\psi$. Each $k^{th}$ antenna introduces some phase delay $\alpha^{k-1}_x = \exp(-j2 \pi (k-1) \frac{h}{\lambda} \sin(\psi_x))$ to the received signal, where $\lambda$ is the wavelength. The index $x=e$ for direction of arrival (DoA) of the eNB $(\psi_e)$ w.r.t $l^{th}$ UE or $x=q$ for DoA of the $q^{th}$ UE $(q \neq l)$ w.r.t $l^{th}$ UE, whose uplink signal results in CCI at downlink of the $l^{th}$ UE. Algorithms like Root-MUSIC (due less computational complexity) can be used for estimating the DoAs. \textit{However, the proposed method is independent of availability of information about DoA of interfering UEs.}\par

     At the $l^{th}$ UE, the downlink received signal at $k^{th}$ antenna is obtained after multiplying the factor $\bigtriangleup_k$ followed by SIC cancellation in receive chain, where $\bigtriangleup_k = (\alpha^{k-1}_e)^*$. This is required for \textit{co-phasing} of the downlink signal from the eNB by removing the phase $(\alpha^{k-1}_e)$ introduced by the $k^{th}$ antenna \cite{mrc}. Hence, the received signal at $k^{th}$ antenna, is given by:   \vspace{-1.00em}
     
     \begin{multline}
       \label{five}
 \mathbf{y}^d_{l,k}= \sum_{j=1}^{N_e} \mathbf{h}^d_{j,l,k} \otimes \mathbf{s}^d_j + \mathbf{n}_{l,k}  \\
+ \Big [ \sum_{\substack{ q=l \\ q \neq l}}^{K} \sum_{\acute{k}=1}^{N_r} \mathbf{h}^u_{q,\acute{k},l,k} \otimes \mathbf{s}^u_{q,\acute{k}}] \alpha^{k-1}_{q}  \bigtriangleup_k 
       \end{multline}

where $ l=1,2,...,K$, $k=1,2,...,N_r$. $\otimes$  denotes the circular convolution operation. $\mathbf{y}^d_{l,k}$ is $N_{cp}X1$ vector where $N_{cp}$ is the receive symbol size with CP. $\mathbf{h}^{d}_{j,l,k}=[h^d_{j,l,k}(0),h^d_{j,l,k}(1),...,h^d_{j,l,k}(L-1),(N_{cp}-L)zeros]^T$ is the complex i.i.d Rayleigh fading channel coefficient between  $j^{th}$ antenna of the eNB and $k^{th}$ antenna of the $l^{th}$ UE in the downlink.  $\mathbf{n}_{l,k} \in N(0,\sigma_n^2 \mathbf{I}_{N_{cp}})$ is the channel noise at $k^{th}$ antenna of the $l^{th}$ UE. $\mathbf{h}^u_{q,\acute{k},l,k}=[h^u_{q,\acute{k},l,k}(0),h^u_{q,\acute{k},l,k}(1),...,h^u_{q,\acute{k},l,k}(L-1)),(N_{cp}-L)zeros]^T$ is the complex i.i.d Rayleigh fading channel coefficient between $\acute{k}^{th}$ antenna of the $q^{th}$ UE and $k^{th}$ antenna of the $l^{th}$ UE. $\mathbf{s}^u_{q,\acute{k}}$ is the uplink signal from $\acute{k}^{th}$ antenna of the $q^{th}$ UE. $\alpha^{k-1}_{q}$ is the spatial response (or the phase delay) of $k^{th}$ antenna of the $l^{th}$ UE in the DoA of $\psi_{q}$. This signal can be represented by:

 \begin{equation}
       \label{six}
 \mathbf{y}^d_{l,k}= \sum_{j=1}^{N_e} \mathbf{h}^d_{j,l,k} \otimes \mathbf{s}^d_j + \mathbf{I}_{l,k} 
       \end{equation} 
       
where $\mathbf{I}_{l,k} = \mathbf{n}_{l,k} + \Big [ \sum_{\substack{ q=l \\ q \neq l}}^{K} \sum_{\acute{k}=1}^{N_r} \mathbf{h}^u_{q,\acute{k},l,k} \otimes \mathbf{s}^u_{q,\acute{k}} \Big ] \alpha^{k-1}_{q}  \bigtriangleup_k  $ is the noise and interference suffered by the $l^{th}$ UE at the $k^{th}$ antenna. After the removal of CP, the signal is converted to the frequency domain:

       \begin{equation}
       \label{seven}
      \mathbf{\tilde{y}}^d_{l,k}= \mathbf{F}_N \mathbf{y}^d_{l,k}  
      =\sum_{j=1}^{N_e} \mathbf{{H}}^d_{j,l,k} \mathbf{e}^d_j + \mathbf{\tilde{I}}_{l,k}  
       \end{equation}
       
where $\mathbf{F}_N$ is the N-point DFT matrix. ${\mathbf{H}}^d_{j,l,k} = diag(\mathbf{F}_N \mathbf{h}^d_{j,l,k})$ is the $N X N$ diagonal matrix whose diagonal elements are frequency domain coefficients between $j^{th}$ transmit antenna of the eNB and $k^{th}$ antenna of the $l^{th}$ UE. 

The signal is then subjected to the subcarrier deallocation and after simplifications \cite{ants,sc_fdma}, we get:

   \begin{equation}
       \label{eight}
     \mathbf{\bar{y}}^d_l (m) =  \mathbf{H}^d_l (m) \sum_{i=1}^{K} \mathbf{P}^i_m \mathbf{\tilde{x}}^d_i (m) + \mathbf{\bar{I}}_l (m)
       \end{equation}

where $\mathbf{\bar{y}}_l (m) = [\bar{y}_{l,1}(m), \bar{y}_{l,2}(m),...,\bar{y}_{l,N_r}(m)]^T$, $\bar{y}_{l,k}(m)$ is the $m^{th}$ element of $\mathbf{\bar{y}}_{l,k} = \mathbf{\bar{A}}^i \mathbf{\tilde{y}}^d_{l,k}$. $\mathbf{\bar{A}}^i$ is the $M X N$ deallocation matrix given by $\mathbf{\bar{A}}^i = (\mathbf{A}^i)^T$. $\mathbf{H}^d_l (m)$  is the $N_r X N_e$ frequency domain channel coefficient matrix of the $l^{th}$ UE on the $m^{th}$ subcarrier. $(k,j)^{th}$ entry in the $\mathbf{H}^d_l (m)$ is the $m^{th}$ diagonal element of matrix $\mathbf{{H}}^d_{j,l,k}$. $\mathbf{\bar{I}}_l (m)$  is channel noise and interference for the $l^{th}$ UE on the $m^{th}$ subcarrier. 

The SVD decomposition of channel matrix  is given by $\mathbf{H}^d_l (m)= \mathbf{U}^d_{m,l} \mathbf{E}_{m,l}^d (\mathbf{V}^d_{m,l})^H$. $\mathbf{U}^d_{m,l}$ is a $N_r X Q$ unitary matrix containing the eigenvectors corresponding to non-zero eigenvalues of $\mathbf{H}^d_l(m) (\mathbf{H}^d_l(m))^H$, $\mathbf{E}^d_{m,l}$  is a $Q X Q$ diagonal matrix containing the non-zero eigenvalues $(\lambda^d_{m,l})$ of $\mathbf{H}^d_l(m) (\mathbf{H}^d_l(m))^H$ such that $\mathbf{E}^d_{m,l}=diag((\lambda^d_{m,l,1})^{1/2},(\lambda^d_{m,l,2})^{1/2}, ...,(\lambda^d_{m,l,Q})^{1/2})$ and $\mathbf{V}^d_{m,l}$ is a $N_e X Q$  unitary matrix containing the eigenvectors corresponding to non-zero eigenvalues of $(\mathbf{H}^d_l(m))^H \mathbf{H}^d_l(m)$. The received signal vector on the $m^{th}$ subcarrier due to all UEs sharing the subcarriers is hence can be given by: 

                          \begin{equation}
       \label{nine}
      \mathbf{\bar{y}}^d (m) = \mathbf{U}^d_m \mathbf{E}^d_m (\mathbf{V}^d_m)^H \mathbf{P}_m \mathbf{\tilde{x}}^d(m) + \mathbf{\bar{I}} (m)
       \end{equation} 
       
where $\mathbf{\bar{y}}^d (m) =  [(\mathbf{\bar{y}}^d_l(m))^T,(\mathbf{\bar{y}}^d_2(m))^T,...,(\mathbf{\bar{y}}^d_K(m))^T]^T$, $\mathbf{U}^d_m= diag(\mathbf{U}^d_{m,1},\mathbf{U}^d_{m,2},...,\mathbf{U}^d_{m,K})$, $\mathbf{V}^d_m = [\mathbf{V}^d_{m,1},\mathbf{V}^d_{m,2},...,\mathbf{V}^d_{m,K}]$, $\mathbf{E}^d_m=diag(\mathbf{E}^d_{m,1}, \mathbf{E}^d_{m,2},...,\mathbf{E}^d_{m,K})$, $\mathbf{P}_m=[\mathbf{P}^1_m, \mathbf{P}^2_m,...,\mathbf{P}^K_m]$, $\mathbf{\tilde{x}}^d(m)=[(\mathbf{\tilde{x}}_1^d(m))^T, (\mathbf{\tilde{x}}_2^d(m))^T,...,(\mathbf{\tilde{x}}_K^d(m))^T]^T$ and $\mathbf{\bar{I}}(m)=[(\mathbf{\bar{I}}_1(m))^T,(\mathbf{\bar{I}}_2(m))^T,...,(\mathbf{\bar{I}}_K(m))^T]^T$. \par

The interference from the downlink of other UEs on the $l^{th}$ UE can be completely eliminated by choosing the precoding matrix as $\mathbf{P}_m=[(\mathbf{V}^d_m)^H]^+ $ which is the pseudo inverse of $\mathbf{V}^d_m$. At the $l^{th}$ UE, the received signal on the $m^{th}$ subcarrier after post-processing, .i.e, multiplying with $(\mathbf{U}^d_{m,l})^H$ is given by \cite{sc_fdma}: \vspace{-0.50em}
      
 \begin{equation}
       \label{ten}
             \mathbf{\tilde{y}}^d_{l} (m) = \mathbf{{E}}^d_{m,l} \mathbf{\tilde{x}}^d_{l} (m) + \mathbf{\tilde{I}}_{l}(m)
        \end{equation} 
       
Now, defining $\mathbf{\bar{E}}^d_{m,l} = [E^d_{m,l}(1,1),E^d_{m,l}(2,2),...,E^d_{m,l}(j,j),...,E^d_{m,l}(Q,Q)]^T$ where $E^d_{m,l}(j,j)$ is the $j^{th}$ diagonal element of $\mathbf{E}^d_{m,l}$. As $\mathbf{\tilde{x}}^d_l(m) = (\bar{x}^d_{l}(m) \mathbf{J}_{1,Q})^T$, above equation can be rearranged to: \vspace{-0.50em}

\begin{equation}
       \label{eleven}
\mathbf{\tilde{y}}^d_{l} (m) = \mathbf{\bar{E}}^d_{m,l} \bar{x}^d_{l} (m) + \mathbf{\tilde{I}}_{l}(m) 
\end{equation}

Now, assuming rich scattering environment and equal channel conditions between the $l^{th}$ UE and $K-1$ interfering UEs (especially in a small cell scenario), the received signal for the $l^{th}$ user on the $m^{th}$ subcarrier can be obtained by MRC \footnote{MMSE equalizer requires knowledge of covariance of the interference at the receiver. As this information is not available at the UE in downlink, MRC is deployed which requires only the downlink channel estimate.} \cite{mrc} as:

   \begin{equation}
       \label{twelve}
      \hat{\bar{x}}^d_l(m)=((\mathbf{\bar{E}}^d_l(m))^H (\mathbf{\bar{E}}^d_l(m))^{-1} (\mathbf{\bar{E}}^d_l(m))^H \mathbf{\tilde{y}}^d_l(m)
       \end{equation} 
       
This signal for the $l^{th}$ UE is then converted to time domain by an M-point IDFT operation given by:

\begin{equation}
       \label{thirteen}
     \mathbf{\hat{x}}^d_l = \mathbf{\bar{F}}_M \mathbf{\hat{\bar{x}}}^d_l
       \end{equation} 
       
where $l=1,2,...,K$ and $\mathbf{\bar{F}}_M$  is M-point inverse IDFT matrix and $\mathbf{\hat{\bar{x}}}^d_l =[\hat{\bar{x}}^d_l (1), \hat{\bar{x}}^d_l (2), ...,\hat{\bar{x}}^d_l (M)]^T$ . This is used for decoding of signal for the $l^{th}$  UE. For the analysis of system performance, the effective SINR $(\gamma_{eff^m_l})$ \footnote{In this work, equal power allocation across all the $Q$ streams and $N_r$ streams is considered at the downlink and uplink, respectively.} is considered for the $l^{th}$ UE on the $m^{th}$ subcarrier from the equation (\ref{twelve}):

\begin{equation}
       \label{fourteen}
      \gamma_{eff^m_l} = [\sigma_{\bar{n}_l}^2 + \sum_{\substack{ q=l \\ q \neq l}}^{K} \sigma_q^2 \beta^q_m ]^{-1}||\mathbf{\bar{E}}^d_{m,l}||^2 (\beta^l_m Q^{-1})
       \end{equation}

where $\sigma_{\bar{n}_l}^2 = E[|\bar{n}_{l,k}(m)|^2], \forall m, l, k$ such that $\bar{n}_{l,k}(m)$ is the frequency domain i.i.d noise variable for $k^{th}$ antenna of the $l^{th}$ UE on the $m^{th}$ subcarrier. $\sum_{\substack{ q=l \\ q \neq l}}^{K} \sigma_q^2 \beta_m^q$ is the total interference power, such that $\sigma_q^2 = |\mathbf{H}^u_{q,\acute{k},l,k}(m)|^2, \forall m,\acute{k}, k, l, k$ where $\mathbf{H}^u_{q,\acute{k},l,k}(m)$ is the frequency domain i.i.d channel coefficient between $\acute{k}^{th}$ antenna of the $q^{th}$ UE and $k^{th}$ antenna of the $l^{th}$ UE on the $m^{th}$ subcarrier and $\beta_m^q = |\tilde{x}^u_q(m)|^2, \forall m$ is the total uplink power allocated to the signal of the $q^{th}$ UE on the $m^{th}$ subcarrier.  $\beta_m^l = |\tilde{x}^d_l(m)|^2, \forall m, l$ is the total power allocated to downlink signal of the $l^{th}$ UE on the $m^{th}$ subcarrier. Now, considering $\beta_m^l \geq \beta_m^q, \forall m,q$, to simplify the equation (\ref{fourteen}), we take $\beta_m^q = \beta_m^u, \forall m,q$, $\beta_m^l= Q\beta_m^u $ and $\sigma_q^2 = \sigma_u^2, \forall q$. Hence, the equation (\ref{fourteen}) can be expressed as: \vspace{-0.50em}

\begin{equation}
       \label{fifteen}
      \gamma_{eff^m_l} = [\sigma_{\bar{n}_l}^2 + (K-1) \sigma_u^2 \beta^u_m ]^{-1}||\mathbf{\bar{E}}^d_{m,l}||^2 \beta_m^u
       \end{equation}

The derivation of the expression is included in the appendix. It is important to observe that the effective SINR can be controlled by the gain factor, .i.e, $||\mathbf{\bar{E}}^d_{m,l}||^2$. This is discussed through simulation results in the next section. 

\section{Simulation Results}
To validate the inclusion of SIC design in the FD system in the proposed architecture, MATLAB simulations are carried out for BER performance in downlink and uplink in \cite{ants}. Here, an FD eNB and three spatially uncorrelated FD UEs (say UE1, UE2 and UE3) sharing the same spectrum resources at both the downlink and uplink, are considered. The capacity \footnote{The capacity here is defined as number of correct bits received per second per Hertz} $(bits/s/Hz)$ at the downlink of UE1 is analyzed with two other UEs acting as the interference source. This work assumes perfect cancellation of self-interference for the FD operation. The channel between each antenna of eNB and UEs' is taken as frequency selective with $L=10$. The modulation scheme used is 16-QAM (no coding). The bandwidth allocated to the UEs is taken to be 3 MHz which is split into 256 subcarriers, out of which 180 subcarriers are occupied by the UEs. A cyclic prefix of duration $4.69 \mu s$ is used. The transmit powers of the UEs are normalized to unity, .i.e, $\beta_m^u = 1, \forall m$ \par

\begin{figure}
\centering
\includegraphics[width=2.75in ,height=1.65in]{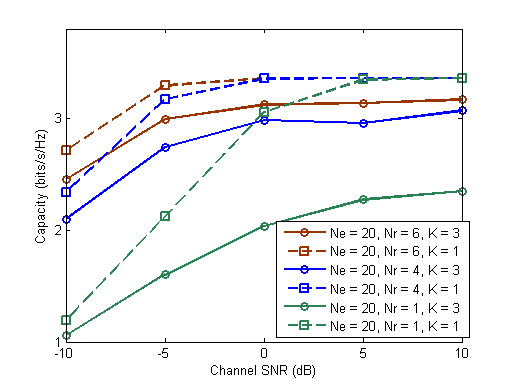}
  % where an .eps filename suffix will be assumed under latex, 
% and a .pdf suffix will be assumed for pdflatex; or what has been declared
% via \DeclareGraphicsExtensions.
\vspace{-0.5em}
   \caption{Downlink capacity vs channel SNR of UE1 for different $N_r$}
  % \vspace{-0.55em}
   \label{fthree}
   \vspace{-1.0em}
\end{figure}

\begin{figure}
\centering
\includegraphics[width=2.75in ,height=1.65in]{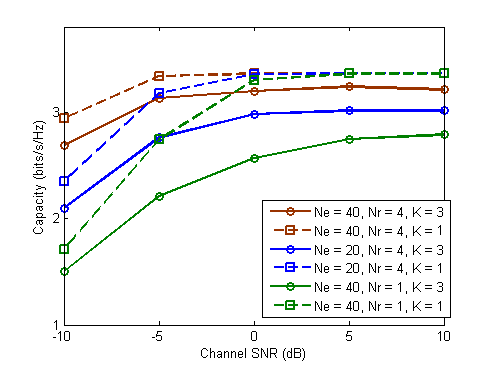}
  % where an .eps filename suffix will be assumed under latex, 
% and a .pdf suffix will be assumed for pdflatex; or what has been declared
% via \DeclareGraphicsExtensions.
\vspace{-0.5em}
   \caption{Downlink capacity vs channel SNR of UE1 for different $N_e$}
   \vspace{-0.45em}
   \label{ffour}
      \vspace{-1.25em}
\end{figure}

Fig.\ref{fthree} and fig.\ref{ffour} show the downlink capacity of UE1 vs. the channel SNR for the increasing number of the receive antennas $(N_r)$ and transmit antennas $(N_e)$ at the UEs and the eNB, respectively. In case of the increasing $N_r$ at the UEs, the number of transmit antennas at the eNB is kept constant at $N_e = 20$. It can be seen from fig.\ref{fthree} that with the increase in $N_r$, there is an improvement in the downlink capacity of the UE. This improvement is due to the increase in the diversity order which results in higher magnitude of $||\mathbf{\bar{E}}^d_{m,l}||^2, \forall m, l$ improving the  $\gamma_{eff^m_l}, \forall m$. Similarly, for the increasing $(N_e)$ at the eNB, the number of receive antennas at the UE is kept constant at $N_r = 4$. An improvement in the downlink capacity of the UE can be observed (fig.\ref{ffour}) with the increase in $N_e$. This is due to increase in magnitude of each eigenvalues, .i.e, $E^d_{m,l}(j,j), j=1,2,...,Q, \forall m, l$ resulting in the higher $||\mathbf{\bar{E}}^d_{m,l}||^2$ and hence improving the $\gamma_{eff^m_l}$. Also, from fig.3 and fig.4, it can be observed that at the higher channel SNR region, the downlink system performance is only interference limited. \par 

In all the above simulations, with the increase in $N_r$ and $N_e$, the downlink capacity approaches the ideal value where there is no CCI, .i.e, $K = 1$. The results are also compared with the conventional scenario of no diversity gain at the UE, .i.e, $N_r =1$. It can be observed that there is a significant improvement in the downlink capacity with the diversity gain. Hence, increasing both $N_r$ and $N_e$ improves the system performance at the downlink in the presence of CCI. However, increasing $N_r$ at the UEs, increases the computational complexity at the UEs including the power consumption. With the power consumption not a constraint at the eNB, comparatively, and profound research on massive MIMO (deploying large $N_e$) in recent years \cite{massive}, the prospect of using large antennas arrays at eNB to tackle the CCI in case of the FD downlink seems so be encouraging. Moreover, the channel reciprocity property of the FD communication will aid the CSI actualization at the eNB.   \par 

\section{Conclusion}
 In this paper, an analysis is carried out to show the use of diversity gain in making the multiuser full-duplex communication feasible. An architecture for the FD eNB and FD UE enabling the corresponding downlink operation is described and performance of the system is studied in terms of the downlink capacity for a UE. From the simulation, it can be seen that the downlink capacity of a UE is significantly improved in the presence of CCI by increasing the number of receive and transmit antennas. However, the use of large number of antennas at eNB is suggested to be a favorable option keeping in mind the power constraint at the UEs and encouraging progress made in the field of massive MIMO.

\section*{Appendix}

The noise and interference in the equation (\ref{six}) is given by:

\begin{equation}
\label{sixteen}
 \mathbf{I}_{l,k} = \mathbf{n}_{l,k} + \Big [ \sum_{\substack{ q=l \\ q \neq l}}^{K} \sum_{\acute{k}=1}^{N_r} \mathbf{h}^u_{q,\acute{k},l,k} \otimes \mathbf{s}^u_{q,\acute{k}} \Big ] \alpha^{k-1}_{q}  \bigtriangleup_k
\end{equation}
       
 After removing the CP and converting it to the frequency domain, we have:
 
 \begin{equation}
\label{seventeen}
\begin{split}
  \mathbf{\tilde{I}}_{l,k} &= \mathbf{F}_N \mathbf{I}_{l,k} \\
   &= \mathbf{\tilde{n}}_{l,k} + \Big [ \sum_{\substack{ q=l \\ q \neq l}}^{K} \sum_{\acute{k}=1}^{N_r} \mathbf{H}^u_{q,\acute{k},l,k} \mathbf{x}^u_{q,\acute{k}} \Big ] \alpha^{k-1}_{q}  \bigtriangleup_k
   \end{split}
\end{equation}

where $\mathbf{H}^u_{q,\acute{k},l,k} = diag(\mathbf{F}_N \mathbf{h}^u_{q,\acute{k},l,k})$ is the $N X N$ diagonal matrix whose diagonal elements are frequency domain coefficients between $\acute{k}^{th}$ transmit antenna of the $q^{th}$ UE and $k^{th}$ receive antenna of the $l^{th}$ UE. $\mathbf{x}^u_{q,\acute{k}}$ is the transmit signal from $\acute{k}$ antenna of the $q^{th}$ UE. This signal is then subjected to the subcarrier deallocation and after simplifications, similar to equation (\ref{eight}), we get:

   \begin{equation}
       \label{eighteen}
     \mathbf{\bar{I}}_l (m) = \mathbf{\bar{n}}_l(m) + \sum_{\substack{ q=l \\ q \neq l}}^{K} \mathbf{\alpha}_q \mathbf{\bigtriangleup} \mathbf{H}^u_{q,l}(m)  \mathbf{\bar{x}}^u_q (m)
       \end{equation}

where $\mathbf{\bar{I}}_l (m) = [\bar{I}_{l,1}(m), \bar{I}_{l,2}(m),... ,\bar{I}_{l,N_r}(m)]^T$, $ \mathbf{\alpha}_q =  diag(\alpha_q^{0}, \alpha_q^{1}, ...., \alpha_q^{N_r-1})$ and $\mathbf{\bigtriangleup} = diag(\bigtriangleup_1, \bigtriangleup_2, ..., \bigtriangleup_{N_r})$. $\mathbf{H}^u_{q,l} (m)$  is the $N_r X N_r$ frequency domain channel coefficient vector between the $l^{th}$ U and the $q^{th}$ UE on the $m^{th}$ subcarrier. This signal for $k^{th}$ antenna of the $l^{th}$ UE is given by:

   \begin{equation}
       \label{nineteen}
     \bar{I}_{l,k}(m) = \bar{n}_{l,k}(m) + \Big [ \sum_{\substack{ q=l \\ q \neq l}}^{K} \sum_{\acute{k}=1}^{N_r} H^u_{q,\acute{k},l,k}(m) \bar{x}^u_{q,\acute{k}}(m) \Big ] \alpha^{k-1}_{q}  \bigtriangleup_k
    \end{equation}
    
Now, considering the fact that the $\bar{n}_{l,k}(m)$ and $H^u_{q,\acute{k},l,k}(m)$ are i.i.d, the power in the above signal is given by:

\begin{equation}
       \label{twenty}
       \begin{split}
     E[|\bar{I}_{l,k}(m)|^2] &= \sigma^2_{\bar{n}_{l}(m)} +  \sum_{\substack{ q=l \\ q \neq l}}^{K} \sum_{\acute{k}=1}^{N_r} \sigma_q^2 (\beta^q_m N_r^{-1}) \\
     & = \sigma^2_{\bar{n}_{l}(m)} + \sum_{\substack{ q=l \\ q \neq l}}^{K} \sigma_q^2 \beta^q_m, \forall l, k
     \end{split}
     \end{equation}
     
The signal is now subjected to post-processing from equation (\ref{ten}), .i.e, multiplying with $(\mathbf{U}^d_{m,l})^H$:

\begin{equation}
\label{twenty_one}
  \mathbf{\tilde{I}}_l(m) = (\mathbf{U}^d_{m,l})^H \mathbf{\bar{I}}_l(m)
\end{equation}
   
where $\mathbf{\tilde{I}}_l(m)=[{\tilde{I}}_{l,1}(m), {\tilde{I}}_{l,2}(m), ...,{\tilde{I}}_{l,Q}(m)]^T$. As $Q=N_r$ and $\mathbf{U}^d_{m,l}$ is an unitary matrix, we have $|\mathbf{\tilde{I}}_l(m)|^2 = |\mathbf{\bar{I}}_l(m)|^2$. The power of the signal at the $j^{th}$ data stream of the $l^{th}$ UE is given by:

\begin{equation}
\label{twenty_two}
\begin{split}
E[|\tilde{I}_{l,j}(m)|^2] &= E[|\bar{I}_{l,k}(m)|^2] \\ 
&= \sigma^2_{\bar{n}_{l}(m)} + \sum_{\substack{ q=l \\ q \neq l}}^{K} \sigma_q^2 \beta^q_m, \forall j
\end{split}
     \end{equation}
     
The received signal for the $l^{th}$ user at the $m^{th}$ subcarrier obtained by MRC is given in equation (\ref{twelve}) as:
   
   \begin{equation}
       \label{twenty_three}
       \begin{split}
      \hat{\bar{x}}^d_l(m) &=((\mathbf{\bar{E}}^d_l(m))^H (\mathbf{\bar{E}}^d_l(m))^{-1} (\mathbf{\bar{E}}^d_l(m))^H \mathbf{\tilde{y}}^d_l(m) \\
        &= \bar{x}_l^d(m) + ((\mathbf{\bar{E}}^d_l(m))^H (\mathbf{\bar{E}}^d_l(m))^{-1} (\mathbf{\bar{E}}^d_l(m))^H \mathbf{\tilde{I}}_l(m)
        \end{split}
       \end{equation} 
       
From this signal, the interference and noise term is represented by:

\begin{equation}
\label{twenty_four}
  I_n = ((\mathbf{\bar{E}}^d_l(m))^H (\mathbf{\bar{E}}^d_l(m))^{-1} (\mathbf{\bar{E}}^d_l(m))^H \mathbf{\tilde{I}}_l(m)
\end{equation}

As the elements of $\mathbf{\tilde{I}}_l(m)$ are i.i.d, the power in the above signal is given by:

\begin{equation}
\label{twenty_five}
 \begin{split}
 E[|I_n|^2] &= [||\mathbf{E}_{m,l}^d||^2]^{-1} |\tilde{I}_{l,k}(m)|^2 \\ 
   & = [||\mathbf{E}_{m,l}^d||^2]^{-1} (\sigma^2_{\bar{n}_{l}(m)} + \sum_{\substack{ q=l \\ q \neq l}}^{K} \sigma_q^2 \beta^q_m) 
   \end{split}
\end{equation}

Hence, the effective SINR $(\gamma_{eff^m_l})$ for the $l^{th}$ UE on the $m^{th}$ subcarrier can now be given by:
\begin{equation}
       \label{twenty_six}
       \begin{split}
      \gamma_{eff^m_l} &= [\sigma_{\bar{n}_l}^2 + \sum_{\substack{ q=l \\ q \neq l}}^{K} \sigma_q^2 \beta^q_m ]^{-1}||\mathbf{\bar{E}}^d_{m,l}||^2 |\bar{x}_l^d (m)|^2 \\
            &= [\sigma_{\bar{n}_l}^2 + \sum_{\substack{ q=l \\ q \neq l}}^{K} \sigma_q^2 \beta^q_m ]^{-1}||\mathbf{\bar{E}}^d_{m,l}||^2 (\beta^l_m Q^{-1})
      \end{split}
       \end{equation}

\ifCLASSOPTIONcaptionsoff
  \newpage
\fi

\end{document}